\documentclass[letterpaper]{article} 
\usepackage[preprint]{aaai2027}  
\usepackage[hyphens]{url}  
\usepackage{graphicx} 
\def\UrlFont{\rm}  
\usepackage{natbib}  
\usepackage{caption} 
\usepackage{algorithm}
\usepackage{algorithmic}

\usepackage{amsmath}
\usepackage{amssymb}

\usepackage{multirow}
\usepackage[table]{xcolor}

\usepackage{newfloat}
\usepackage{listings}
\DeclareCaptionStyle{ruled}{labelfont=normalfont,labelsep=colon,strut=off} 
\floatstyle{ruled}
\newfloat{listing}{tb}{lst}{}
\floatname{listing}{Listing}

\usepackage{booktabs}

\title{MedDiT4SR: Tri-Stream Joint Adaptation of Pre-Trained Diffusion Transformers for Medical Image Super-Resolution}
\author{
    Zhi Chen\textsuperscript{\rm 1},
    Le Zhang\textsuperscript{\rm 1}\corresponding
}

\affiliations{
    \textsuperscript{\rm 1}School of Engineering, University of Birmingham\\
    zxc561@student.bham.ac.uk,
    l.zhang.16@bham.ac.uk
}

\begin{document}

\maketitle

\begin{abstract}
Medical image super-resolution (MedSR) requires recovering fine anatomical structures from degraded observations while avoiding unsupported details introduced by generative priors. Large-scale pre-trained multimodal diffusion transformers provide strong visual priors, but their adaptation to MedSR remains non-trivial. In conventional ControlNet-style adaptation, the low-resolution (LR) image is processed as an external condition and injected into the denoising stream through one-way connections. Consequently, LR anatomical evidence cannot be jointly updated with the evolving denoising and semantic representations.
We propose \textbf{MedDiT4SR}, a tri-stream adaptation framework that integrates the LR, noisy latent, and text representations into the same multimodal diffusion-transformer blocks. To complement global token interaction, we introduce a \textbf{Super-Resolution Adapter (SR Adapter)} that aggregates scale-dependent local tokens and suppresses interpolation-induced redundancy. We further propose a \textbf{Semantic Alignment Refiner (SA Refiner)} that calibrates local LR responses using prompt-conditioned semantic information.
Experiments under both in-domain and within-modality cross-dataset settings demonstrate the effectiveness of adapting large-scale pre-trained DiT models to medical image super-resolution across diverse imaging domains.
\end{abstract}


\begin{links}
    \link{Code}{https://github.com/zrs03/MedDiT4SR}
\end{links}

\section{Introduction}

\begin{figure}[t]
\centering
\includegraphics[width=0.98\columnwidth]{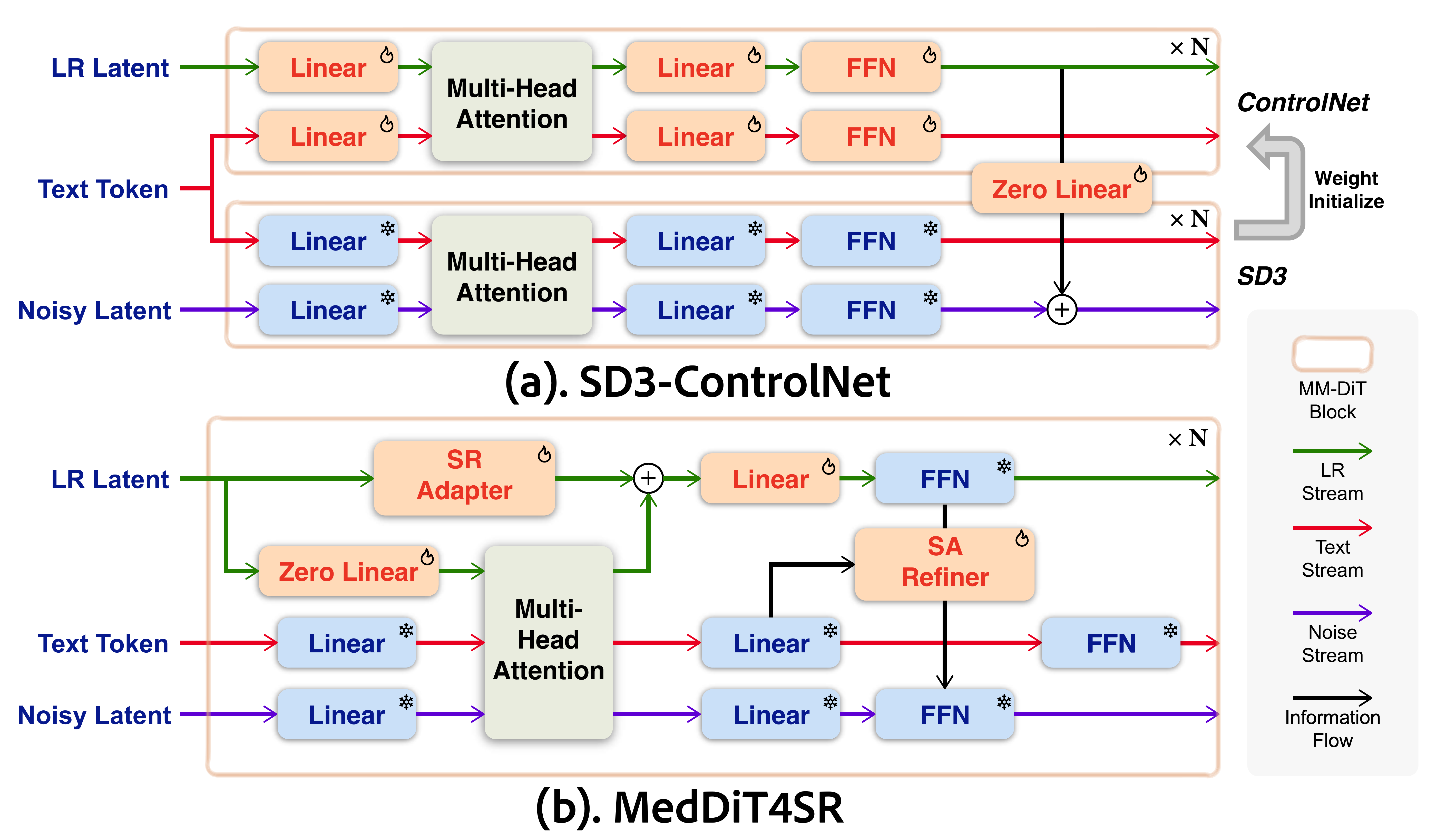}
\caption{SD3-ControlNet processes LR features in a separate branch and injects them into the Noise Stream through one-way connections. MedDiT4SR instead integrates the LR Stream into joint attention, jointly updating the LR, Noise, and Text Streams.}
\label{fig:comparison}
\end{figure}

Medical image super-resolution (MedSR) aims to recover a high-resolution (HR) medical image from its low-resolution (LR) counterpart. Unlike natural image super-resolution, MedSR requires the model not only to restore fine anatomical structures from degraded observations but also to faithfully preserve clinically relevant details without introducing hallucinated artifacts, as inaccurate reconstruction may compromise subsequent clinical diagnosis and analysis. Therefore, MedSR has attracted increasing attention as an effective solution for enhancing medical image quality.
With the remarkable advances in deep learning, a variety of neural network-based methods have been proposed for MedSR, ranging from traditional convolutional neural networks (CNNs)~\cite{liu2019medicalimagesuperresolutionmethod,zhao2020smore} to the recently emerging diffusion-based models~\cite{tu2025taming}.

Recently, Diffusion Transformers (DiTs) have emerged as a powerful alternative to conventional UNet-based diffusion models. Benefiting from the remarkable success of large-scale pre-trained models such as SD3~\cite{esser2024scalingrectifiedflowtransformers} and FLUX~\cite{flux2024}, which are pre-trained on massive image-text corpora and encode rich real-world visual priors, Multimodal Diffusion Transformers (MM-DiTs) have demonstrated outstanding scalability, image quality, and cross-domain generalization. Unlike conventional diffusion backbones, MM-DiT maintains independent streams for image and text representations while enabling bidirectional information interaction through joint attention.
Such rich visual priors and multimodal interaction capabilities are particularly appealing for MedSR, where faithful anatomical reconstruction often requires both strong generative priors and semantic understanding.
These advances naturally raise an important question: \emph{Can large-scale pre-trained DiT models be effectively adapted for medical image super-resolution?}

A straightforward approach is to adapt ControlNet~\cite{zhang2023addingconditionalcontroltexttoimage} for MM-DiT by injecting LR information into the diffusion process, as illustrated in Figure~\ref{fig:comparison}(a). Specifically, several MM-DiT blocks are duplicated as a trainable branch initialized from the pre-trained backbone. The LR image is encoded by the pre-trained VAE encoder into a latent representation, which is processed by the duplicated blocks. The resulting hidden features are projected through trainable convolution layers and injected into the Noise Stream of SD3. While effective as external conditioning, this one-way feature injection prevents the LR representation from directly participating in the joint attention mechanism of MM-DiT. As a result, degradation-aware structural cues cannot evolve together with the Noise Stream and Text Stream, restricting multimodal information interaction and limiting reconstruction performance.

These limitations reveal two fundamental challenges in adapting MM-DiTs to MedSR. First, although joint attention enables global multimodal interaction, it primarily models long-range dependencies and is less effective at capturing fine-grained local anatomical structures that are critical for faithful reconstruction. Second, medical image super-resolution also requires semantic guidance to remain consistent with anatomical descriptions, yet effectively integrating text semantics with degradation-aware structural information throughout the denoising process remains an open problem.

To leverage the strengths of large-scale pre-trained MM-DiTs, we move beyond the conventional ControlNet-based adaptation paradigm and propose \textbf{MedDiT4SR}, a control architecture designed to be applicable to medical image super-resolution across different imaging modalities.
MedDiT4SR integrates an additional LR Stream into the original MM-DiT blocks, allowing the LR Stream, Noise Stream, and Text Stream to interact through joint attention. This design enables degradation-aware anatomical information to evolve together with the denoising representation and provides progressively refined structural guidance throughout the diffusion process. Furthermore, we introduce a \textbf{Super-Resolution Adapter (SR Adapter)} to enhance local structural representations by filtering interpolation-induced redundancy and injecting structure-aware priors into the LR Stream. To further exploit semantic information from text prompts, we design a \textbf{Semantic Alignment Refiner (SA Refiner)}, which modulates LR features using text-guided diffusion representations. As illustrated in Figure~\ref{fig:comparison}(b), these components facilitate the integration of structural and semantic guidance for medical image super-resolution.

Our main contributions are summarized as follows:

\begin{itemize}
    \item We are the first to investigate the adaptation of large-scale pre-trained diffusion transformers to medical image super-resolution.
    
    \item We propose a Super-Resolution Adapter (SR Adapter) to enhance degradation-aware local structural guidance in the LR Stream.
    
    \item We propose a Semantic Alignment Refiner (SA Refiner) to introduce prompt-conditioned semantic modulation into the super-resolution process.
    
    \item We achieve SOTA performance across different medical imaging modalities under both in-domain and cross-domain settings.
    
\end{itemize}

\section{Related Work}

\subsection{Medical Image Super-Resolution}

Image Super-Resolution (ISR) has been widely studied in computer vision, with architectures evolving from convolutional networks~\cite{dai2019second, dong2014learning} to transformers~\cite{chen2021pre, chen2023activating}. Medical image super-resolution is particularly important because high-resolution medical images can reveal finer anatomical structures and lesion details, while direct high-resolution acquisition is often constrained by scanning time, hardware limitations, radiation dose, motion artifacts, and patient comfort.

Recent studies have explored medical image super-resolution across different modalities. For example, SMORE improves the through-plane resolution of anisotropic MR images through a self-supervised anti-aliasing and super-resolution framework~\cite{zhao2020smore}, while GAN-CIRCLE~\cite{You_2020} introduces a semi-supervised GAN-based framework with identity, residual, and cycle-consistency constraints for CT super-resolution. Beyond these modality-specific methods, GAN-based medical SR has also been investigated to enhance perceptual details by mapping low-resolution medical images to high-resolution counterparts~\cite{ahmad2022new}.
However, due to the scarcity and high cost of paired high-resolution medical data, most existing medical SR methods are trained on limited specific datasets, making them sensitive to data distributions, degradation patterns, scanners, and acquisition protocols. As a result, their performance may degrade on unseen datasets or real-world clinical scenarios. Moreover, hallucinated structures can compromise clinical reliability. Therefore, leveraging large-scale pre-trained priors is a promising direction for building more robust and faithful medical SR frameworks.

\subsection{Diffusion Transformer}

To improve the scalability and generative capacity of diffusion models, transformer-based architectures have recently been introduced as an alternative to conventional U-Net backbones. Among them, the Diffusion Transformer (DiT)~\cite{peebles2023scalablediffusionmodelstransformers} has demonstrated strong potential by formulating the denoising process with transformer blocks.
In particular, SD3~\cite{esser2024scalingrectifiedflowtransformers} adopts Multimodal Diffusion Transformers (MM-DiTs), where text tokens and image latent tokens are jointly modeled through attention, enabling effective cross-modal information interaction. 
Beyond text-to-image generation, the paradigm has recently been extended to various downstream visual tasks, including prompt-based image editing~\cite{shin2025exploringmultimodaldiffusiontransformers}, image restoration~\cite{kong2025dualpromptingimagerestoration}, and real-world image super-resolution~\cite{duan2025dit4srtamingdiffusiontransformer}. 
These studies suggest that multimodal attention in diffusion transformers provides a flexible mechanism for integrating heterogeneous conditions, such as text prompts, degraded images, and low-resolution guidance.
Therefore, it is worthwhile to explore how established transformer-based interaction mechanisms can be effectively integrated into this powerful diffusion backbone for medical image super-resolution. 

\begin{figure*}[t]
\centering
\includegraphics[width=0.95\textwidth]{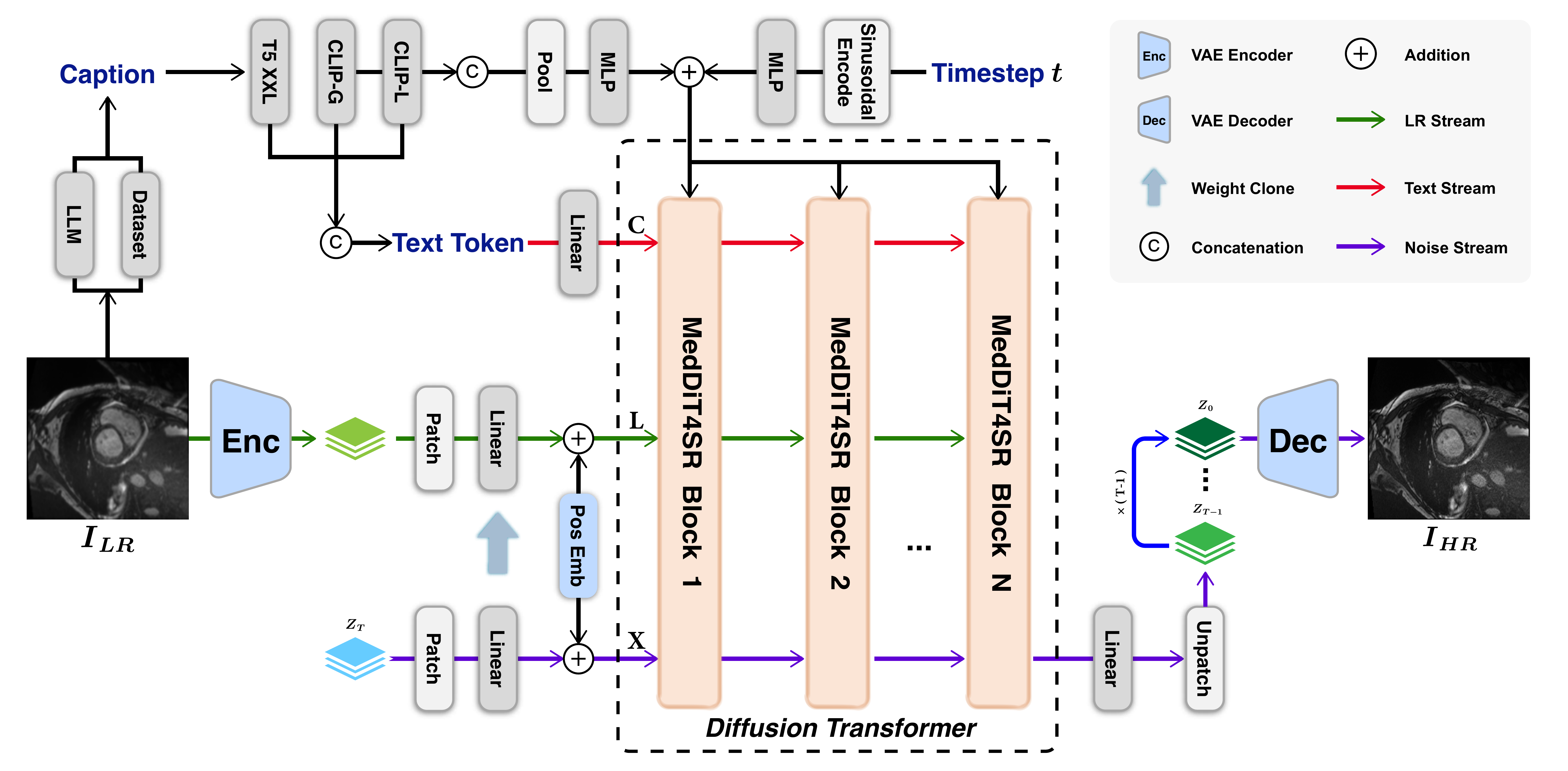}
\caption{Overview of the proposed MedDiT4SR framework for medical image super-resolution.}
\label{fig:overview}
\end{figure*}

\section{Method}
Given a low-resolution medical image $I_{LR}$ affected by complex degradations, MedDiT4SR aims to recover the corresponding high-resolution image $I_{HR}$ while generating realistic and anatomically faithful details. To address the ill-posed nature of medical image super-resolution, we build MedDiT4SR upon the DiT-based SD3 backbone, leveraging the strong generative priors and multimodal interaction capability of diffusion transformers.

\subsection{Overview of Architecture}

Since MedDiT4SR is built upon the widely adopted DiT-architectured SD3, we first briefly introduce SD3. 
Similar to previous SD models~\cite{rombach2022highresolutionimagesynthesislatent}, SD3 performs diffusion in the latent space. 
It consists of a sequence of MM-DiT blocks, where text and image embeddings are processed by modality-specific weights, forming the Text Stream and the Noise Stream, as shown in Figure~\ref{fig:comparison}(a). 
Their token sequences are merged during attention computation, enabling bidirectional cross-modal interaction throughout the denoising process. 
When ControlNet~\cite{von-platen-etal-2022-diffusers, zhang2023addingconditionalcontroltexttoimage} is adopted as a DiT controller, the LR Stream is usually processed by additional DiT blocks and injected into the Noise Stream through trainable linear layers. 
Such a design mainly provides one-way guidance from the LR Stream to the Noise Stream. 
In contrast, MedDiT4SR directly incorporates the LR Stream into the original DiT blocks, enabling the Noise Stream, LR Stream, and Text Stream to interact within the same joint attention operation, as illustrated in Figure~\ref{fig:comparison}(b).

The overall architecture is shown in Figure~\ref{fig:overview}. 
Before entering the diffusion transformer, the noisy latent $\mathbf{Z} \in \mathbb{R}^{H \times W \times C}$ is patchified and projected into the noisy image token $\mathbf{X} \in \mathbb{R}^{K \times D}$, where $K=\frac{H}{2}\cdot\frac{W}{2}$. 
To introduce LR guidance, the LR image $I_{LR}$ is first upsampled to the target spatial resolution by bicubic interpolation, denoted as $I_{LR}^{\mathrm{bic}}$, so that its latent representation has the same spatial size as the noisy latent. 
Then, $I_{LR}^{\mathrm{bic}}$ is encoded by the pre-trained VAE encoder and processed with the same patch and position embedding procedure to obtain the LR image token $\mathbf{L} \in \mathbb{R}^{K \times D}$. 
Following SD3~\cite{esser2024scalingrectifiedflowtransformers}, the caption describing $I_{LR}$ is encoded by CLIP-L~\cite{radford2021learningtransferablevisualmodels}, CLIP-G~\cite{Cherti_2023}, and T5-XXL~\cite{raffel2023exploringlimitstransferlearning}, producing the text token $\mathbf{C} \in \mathbb{R}^{M \times D}$ and pooled text representations for timestep modulation.

Given $\mathbf{X}$, $\mathbf{L}$, and $\mathbf{C}$, each MedDiT4SR block projects them into corresponding query, key, and value representations and concatenates them before joint attention. 
This modification allows LR anatomical guidance to participate directly in the denoising process rather than being injected after attention. 
After $N$ MedDiT4SR blocks and the unpatch operation, the Noise Stream outputs the denoised latent at timestep $t$. 
By repeating the diffusion process for $T$ steps and decoding the clean latent $\hat{\mathbf{Z}}_0$, we obtain the estimated HR result $\hat{I}_{HR}$. 

\begin{figure}[t]
\centering
\includegraphics[width=0.98\columnwidth]{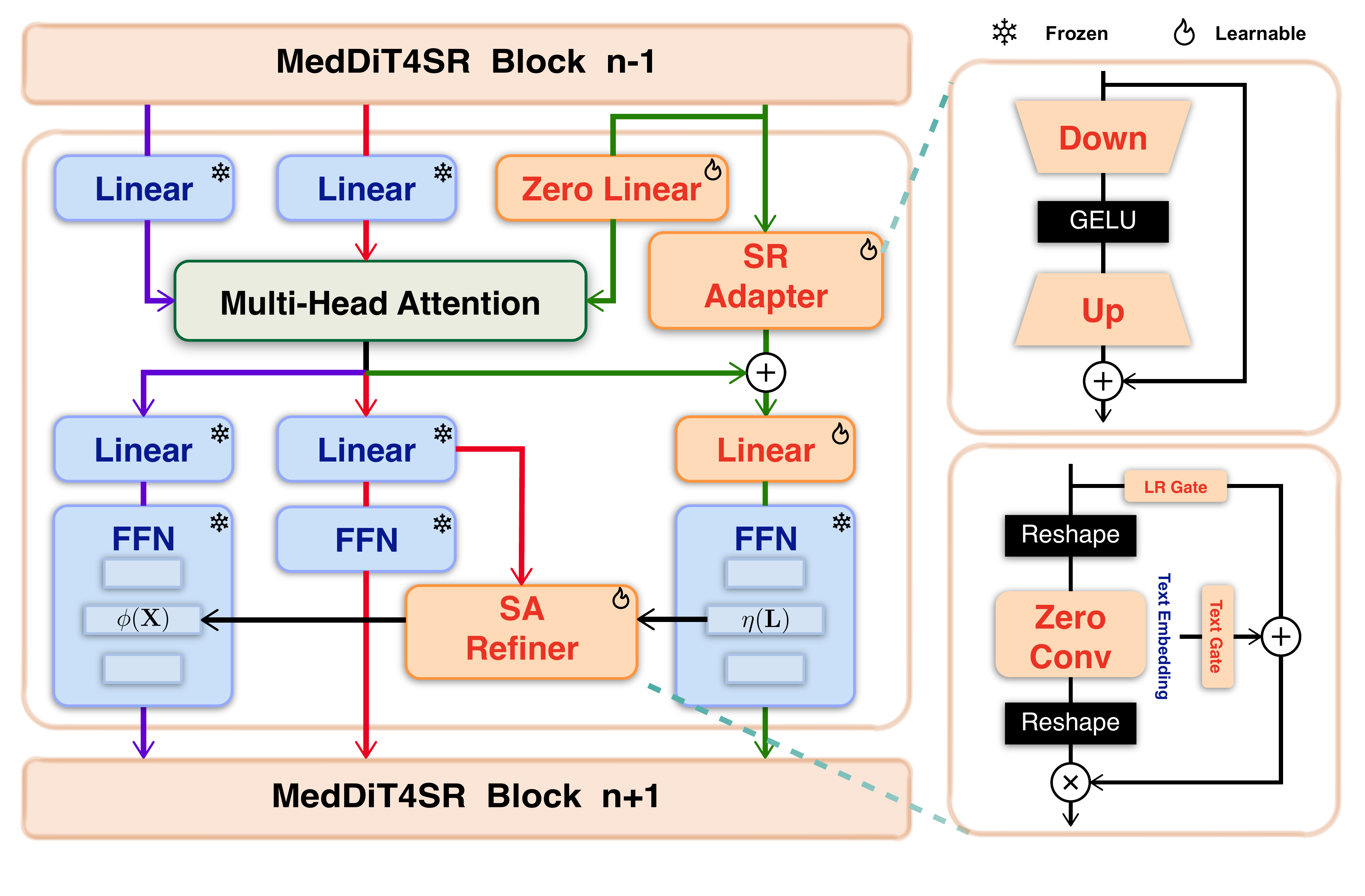}
\caption{MedDiT4SR block architecture.
The LR, Noise, and Text Streams are jointly updated through MM-DiT attention, enabling information exchange among the three streams.
The SR Adapter enhances local structural representations by injecting scale-aware structural priors into the LR Stream, while the SA Refiner adaptively aligns LR features with text semantics via lightweight semantic modulation.}
\label{fig:block}
\end{figure}

\subsection{Structure Guidance with SR Adapter}

Although the LR Stream is incorporated into the joint attention mechanism, attention mainly captures long-range token interactions and may be less effective in modeling local structural patterns. 
Moreover, the bicubic-upsampled LR image inevitably contains redundant and over-smoothed information caused by interpolation. 

To alleviate this issue, we introduce a Super-Resolution Adapter (SR Adapter) to provide structure-aware guidance for the LR Stream. 
The key idea is to aggregate local LR tokens according to the super-resolution scale, filter interpolation-induced redundancy, and inject the refined structural representation back into the LR Stream through a residual connection. 
To ensure stable adaptation from the pre-trained SD3 backbone, the upsampling layer of the SR Adapter is zero-initialized.

As shown in Figure~\ref{fig:block}, given the LR token sequence $\mathbf{L} \in \mathbb{R}^{K \times D}$, we first reshape it into a 2D feature map $\mathbf{F}_{L} \in \mathbb{R}^{h \times w \times D}$, where $K=h \times w$. 
Considering the upscaling factor $s$, inspired by the patch merging operation in Swin Transformer~\cite{liu2021swintransformerhierarchicalvision}, the SR Adapter performs patch-level fusion using a strided convolution with kernel size $s$ and stride $s$. 
This operation groups each non-overlapping $s \times s$ region into a compact representation, enabling the model to suppress redundant interpolated details while retaining scale-aware anatomical structures. 
The fused feature is then activated by GELU and projected back to the original spatial resolution through a transposed convolution. 
This process can be formulated as
\[
\mathbf{F}_{A}
=
\mathrm{Up}_{s}
\left(
\sigma
\left(
\mathrm{Down}_{s}(\mathbf{F}_{L})
\right)
\right),
\]
where $\mathrm{Down}_{s}(\cdot)$ denotes the scale-aware strided convolution, $\mathrm{Up}_{s}(\cdot)$ denotes the transposed convolution, and $\sigma(\cdot)$ is the GELU activation. 
After being reshaped back to the token sequence, the resulting structural residual is added to the original LR token to produce the refined LR representation. 
In this way, the SR Adapter complements the global interaction of joint attention with local structure-aware refinement, helping preserve anatomical details.

\subsection{Semantic Guidance with SA Refiner}

While the SR Adapter enhances local structural representations, semantic alignment with the text prompt remains important. 
Therefore, it is necessary to incorporate the text condition into the LR guidance in a lightweight manner. 
To this end, as illustrated in Figure~\ref{fig:block}, we propose a Semantic Alignment Refiner (SA Refiner) in the feed-forward network (FFN) branch, aiming to achieve prompt-conditioned semantic guidance without explicitly expanding or concatenating prompt features with all LR tokens.

The idea behind SA Refiner is to use the prompt embedding to generate a token-wise modulation gate for the LR features. 
Instead of generating additional prompt tokens or a large set of dynamic parameters, we use lightweight projections to produce scalar guidance scores from the LR feature and the text embedding. 
These scores are then applied to the local LR response, enabling efficient feature-level semantic interaction with negligible extra parameters.

Specifically, we denote the intermediate FFN features of the Noise Stream and LR Stream as $\phi(\mathbf{X}) \in \mathbb{R}^{K \times D_{ff}}$ and $\eta(\mathbf{L}) \in \mathbb{R}^{K \times D_{ff}}$, respectively. 
The SA Refiner conducts semantic refinement over $\eta(\mathbf{L})$. 
We first reshape $\eta(\mathbf{L})$ into a 2D feature map and apply a zero-initialized depth-wise convolution to obtain the local LR response, which can be represented as:
\[
\mathbf{R}_{L} = \mathrm{ZeroConv}(\eta(\mathbf{L})),
\]
where $\mathrm{ZeroConv}(\cdot)$ denotes the zero-initialized depth-wise convolution. 
In the meantime, the text token sequence $\mathbf{C} \in \mathbb{R}^{M \times D}$ is averaged along the token dimension to obtain a compact prompt representation. 
Then, the LR feature and the prompt representation are projected into scalar scores and combined to form a token-wise semantic gate:
\[
\mathbf{G}
=
1
+
\alpha \cdot
\tanh
\left(
\mathbf{W}_{L}\eta(\mathbf{L})
+
\mathbf{W}_{C}\mathrm{Avg}(\mathbf{C})
\right),
\]
where $\mathbf{W}_{L}$ and $\mathbf{W}_{C}$ denote lightweight linear projections, and $\alpha$ is a learnable scale initialized to zero. 
Similar to spatial attention~\cite{woo2018cbamconvolutionalblockattention}, $\mathbf{G}$ can be regarded as a token-level attention map over LR features, which adaptively modulates the contribution of each LR token according to both local image content and prompt-level semantics. 
Finally, the semantic-aware LR response is obtained by applying this gate to the local LR response:
\[
\mathbf{R}_{sem}
=
\mathbf{R}_{L} \odot \mathbf{G}.
\]
The refined feature $\mathbf{R}_{sem}$ is added to $\phi(\mathbf{X})$ within the FFN branch, allowing the Noise Stream to receive prompt-aligned LR guidance. 
In this way, SA Refiner complements the structural guidance of the SR Adapter with semantic-aware modulation for faithful and condition-consistent medical image super-resolution.

\begin{figure*}[t]
\centering
\includegraphics[width=0.99\textwidth]{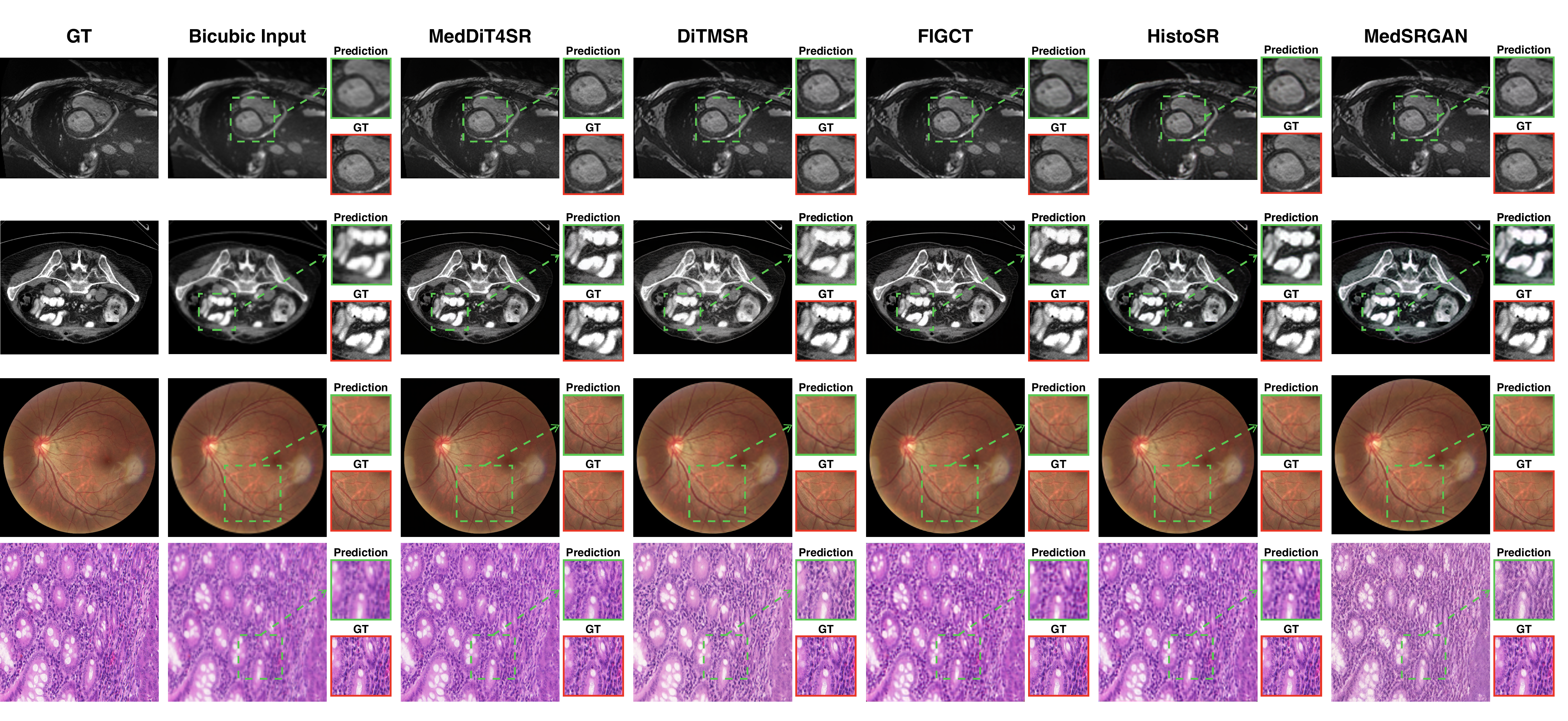}
\caption{Visual comparison of MedDiT4SR with representative baselines presented in Table~\ref{tab:multi} for $\times4$ super-resolution across different imaging modalities. From top to bottom are cardiac MRI, abdominal CT, fundus, and histopathology images.}
\label{fig:vis}
\end{figure*}

\section{Experiments}

\subsection{Datasets}

We conduct experiments on five medical image super-resolution datasets spanning cardiac MRI, abdominal CT, retinal fundus imaging, histopathology microscopy, and breast ultrasound.
For in-domain evaluation, we use five public datasets: ACDC~\cite{bernard2018deep}, BTCV~\cite{landman2015miccai}, REFUGE2~\cite{orlando2020refuge}, GlaS~\cite{sirinukunwattana2017gland}, and BUSI~\cite{al2020dataset}. We train an independent MedDiT4SR model on each dataset. 
For within-modality cross-dataset evaluation, we further use M\&M~\cite{campello2021multi}, AMOS~\cite{ji2022amos}, FLARE22~\cite{ma2024unleashing}, BUSBRA~\cite{gomez2024bus}, and BUSUC~\cite{iqbal2024memory}. In each experiment, a model is trained on a source dataset and directly evaluated on an unseen target dataset from the same imaging modality without target-domain adaptation. 

\subsection{Implementation Details}


The LR--HR pairs are constructed following the preprocessing pipeline of MedSRGAN~\cite{gu2020medsrgan}. All ground-truth HR images are fixed at 512$\times$512. For the $\times$2 and $\times$4 settings, the corresponding LR images are generated by applying scale-specific degradation to the HR images, resulting in resolutions of 256$\times$256 and 128$\times$128, respectively, and are then bicubically upsampled to 512$\times$512 before being fed into the model.
All datasets are divided into training, validation, and test sets with a ratio of $8{:}1{:}1$. 


We train the model using four NVIDIA A100 40GB GPUs, with a batch size of 32 and a constant learning rate of $3\times10^{-5}$. During inference, we adopt the default SD3.5 sampling schedule with 40 sampling steps and set the classifier-free guidance scale to 8. For datasets without accompanying textual descriptions, we follow prior work~\cite{li2025tamingstablediffusioncomputed} and use LLaVA-Med~\cite{li2023llavamedtraininglargelanguageandvision} to generate captions solely from the LR inputs.
Performance is evaluated using PSNR, SSIM, and FID.
Higher PSNR and SSIM values and lower FID indicate better reconstruction quality.
More details about implementation are shown in the supplementary material.

\subsection{Main Results}

\subsubsection{Comparison Across Imaging Modalities}

\begin{table*}[t]
\centering
\setlength{\tabcolsep}{4pt}
\renewcommand{\arraystretch}{1.1}
\caption{Comparison of MedDiT4SR with SOTA  medical image super-resolution methods across multiple imaging modalities for $\times4$ super-resolution. Methods specifically designed for a given modality are highlighted with a grey background.}
\label{tab:multi}

\begin{tabular}{@{}cc ccccccccccc@{}}
\toprule

\multirow{2}{*}{\textbf{Task}} & \multirow{2}{*}{\textbf{Methods}} 
& \multicolumn{3}{c}{\textbf{ACDC}} 
& \multicolumn{2}{c}{\textbf{BTCV}} 
& \multicolumn{2}{c}{\textbf{REFUGE2}} 
& \multicolumn{2}{c}{\textbf{GlaS}} 
& \multicolumn{2}{c}{\textbf{BUSI}} \\

\cmidrule(lr){3-5} \cmidrule(lr){6-7} \cmidrule(lr){8-9} \cmidrule(lr){10-11} \cmidrule(lr){12-13}

& 
& SSIM & PSNR & FID
& SSIM & PSNR
& SSIM & PSNR
& SSIM & PSNR
& SSIM & PSNR \\

\midrule

\multirow{2}{*}{MRI}
& DiTMSR & \cellcolor{gray!15}90.4 & \cellcolor{gray!15} 33.4& \cellcolor{gray!15} 27.3
& 77.1 & 24.6 
& 79.1 & 23.0 
& 78.6 & 24.3 
& 79.2 & 24.7 \\
& PromptMR  & \cellcolor{gray!15}89.7 & \cellcolor{gray!15} 31.9& \cellcolor{gray!15} 31.5
& 74.9 & 22.8 
& 81.8 & 25.4 
& 74.9 & 22.2 
& 77.4 & 24.5 \\

\midrule

\multirow{2}{*}{CT}
& FIGCT & 85.2 & 27.8 & 39.6
& \cellcolor{gray!15}81.6 & \cellcolor{gray!15}29.2 
& 82.6 & 25.7 
& 80.2 & 24.7 
& 70.5 & 23.3 \\
& CT-SRGAN    & 87.5 & 29.9 & 38.2
& \cellcolor{gray!15}80.5 & \cellcolor{gray!15}28.7
& 85.7 & 28.9 
& 81.7 & 25.8 
& 69.8 & 22.4 \\

\midrule

Fundus
& Fundus-GAN & 88.3 & 30.2 & 30.7 
& 80.7 & 28.8 
& \cellcolor{gray!15}86.6 & \cellcolor{gray!15}29.5 
& 82.1 & 26.3 
& 80.7 & 25.3 \\

\midrule

Histopathology
& HistoSR   & 84.9 & 27.4 &  45.2
& 79.0 & 26.7 
& 77.4 & 21.3 
& \cellcolor{gray!15}\textbf{86.3} & \cellcolor{gray!15}28.9 
& 80.1 & 24.9 \\

\midrule

Ultrasound
& DMUISR & 84.7 & 28.0 &  44.2
& 76.2 & 33.8 
& 84.7 & 28.1 
& 74.8 & 21.9 
& \cellcolor{gray!15}86.2 & \cellcolor{gray!15}28.5 \\

\midrule

\multirow{3}{*}{General}
& MedSRGAN   & \cellcolor{gray!15} 89.4 & \cellcolor{gray!15} 30.6 & \cellcolor{gray!15}  28.6
& \cellcolor{gray!15} 81.7 & \cellcolor{gray!15} 29.3 
& \cellcolor{gray!15} 84.3 & \cellcolor{gray!15} 28.4 
& \cellcolor{gray!15} 84.9 & \cellcolor{gray!15} 27.5 
& \cellcolor{gray!15} 85.4 & \cellcolor{gray!15} 28.0 \\

& SG-SRNet   & \cellcolor{gray!15} 86.2 & \cellcolor{gray!15} 28.2 & \cellcolor{gray!15}  32.3
& \cellcolor{gray!15} 75.1 & \cellcolor{gray!15} 23.9 
& \cellcolor{gray!15} 86.2 & \cellcolor{gray!15} 29.1 
& \cellcolor{gray!15} 83.7 & \cellcolor{gray!15} 26.8 
& \cellcolor{gray!15} 82.6 & \cellcolor{gray!15} 26.6 \\

& SD3-ControlNet  & \cellcolor{gray!15} 84.2 & \cellcolor{gray!15} 26.9 & \cellcolor{gray!15} 32.4 
& \cellcolor{gray!15} 79.4 & \cellcolor{gray!15} 26.8 
& \cellcolor{gray!15} 87.6 & \cellcolor{gray!15} 32.9 
& \cellcolor{gray!15} 86.1 & \cellcolor{gray!15} 28.8 
& \cellcolor{gray!15} 83.4 & \cellcolor{gray!15} 27.2 \\

\midrule

\multirow{1}{*}{Proposed}
& \textbf{MedDiT4SR} 
& \cellcolor{gray!15} \textbf{91.2}& \cellcolor{gray!15} \textbf{34.4}& \cellcolor{gray!15}  \textbf{24.4}
& \cellcolor{gray!15} \textbf{83.4} & \cellcolor{gray!15} \textbf{31.0} 
& \cellcolor{gray!15} \textbf{90.3} & \cellcolor{gray!15} \textbf{35.3} 
& \cellcolor{gray!15} 86.0 & \cellcolor{gray!15} \textbf{29.4} 
& \cellcolor{gray!15} \textbf{87.2} & \cellcolor{gray!15} \textbf{29.7} \\

\bottomrule
\end{tabular}

\end{table*}

\begin{table*}[t]
\centering
\setlength{\tabcolsep}{6pt}
\renewcommand{\arraystretch}{1.1}
\caption{Cross-dataset generalization: Models are trained on one dataset and evaluated on another dataset without adaptation. PSNR values are reported under scale factors $\times2$ and $\times4$.}
\label{tab:cross_dataset}
\begin{tabular}{ccccccccccc}
\toprule
\multirow{3}{*}{\textbf{Methods}}
& \multicolumn{4}{c}{\textbf{Breast Ultrasound}}
& \multicolumn{4}{c}{\textbf{Abdominal CT}}
& \multicolumn{2}{c}{\textbf{Cardiac MRI}} \\
\cmidrule(lr){2-5}
\cmidrule(lr){6-9}
\cmidrule(lr){10-11}
& \multicolumn{2}{c}{BUSI$\rightarrow$BUSBRA}
& \multicolumn{2}{c}{BUSI$\rightarrow$BUSUC}
& \multicolumn{2}{c}{BTCV$\rightarrow$AMOS}
& \multicolumn{2}{c}{BTCV$\rightarrow$FLARE-2022}
& \multicolumn{2}{c}{ACDC$\rightarrow$M\&M} \\
\cmidrule(lr){2-3}
\cmidrule(lr){4-5}
\cmidrule(lr){6-7}
\cmidrule(lr){8-9}
\cmidrule(lr){10-11}
& $\times2$ & $\times4$
& $\times2$ & $\times4$
& $\times2$ & $\times4$
& $\times2$ & $\times4$
& $\times2$ & $\times4$ \\
\midrule

DiTMSR
& 26.4 & 23.9
& 25.8 & 23.3
& 26.4 & 24.1
& 26.5 & 24.2
& 32.9 & 31.8 \\

FIGCT
& 25.4 & 22.9
& 25.7 & 23.0
& 30.6 & 28.4
& 29.5 & 27.6
& 29.8 & 27.1 \\

Fundus-GAN
& 27.5 & 24.6
& 27.8 & 24.3
& 29.9 & 28.1
& 30.1 & 27.9
& 30.6 & 28.1 \\

HistoSR
& 26.5 & 24.4
& 26.8 & 24.7
& 28.9 & 26.2
& 29.0 & 26.5
& 30.8 & 26.2 \\

MedSRGAN
& 29.7 & 27.1
& 29.1 & 26.6
& 29.9 & 28.2
& 29.3 & 27.7
& 29.0 & 27.6 \\

SD3-ControlNet
& 29.6 & 26.9
& 29.6 & 26.7
& 30.0 & 27.9
& 29.4 & 27.5
& 32.5 & 30.8 \\

\midrule

\textbf{MedDiT4SR}
& \textbf{30.4} & \textbf{28.6}
& \textbf{30.5} & \textbf{29.0}
& \textbf{31.4} & \textbf{30.5}
& \textbf{31.2} & \textbf{29.6}
& \textbf{34.6} & \textbf{33.1} \\

\bottomrule
\end{tabular}
\end{table*}










\begin{table*}[t]
\centering
\caption{Ablation study of multimodal attention, structure guidance, 
and semantic guidance under a $\times4$ super-resolution setting.}
\label{tab:ablation}
\setlength{\tabcolsep}{7pt}
\renewcommand{\arraystretch}{1.1}

\begin{tabular}{cc cc cc cc cc}
\toprule
\multicolumn{2}{c}{\textbf{Multimodal Attention}}
& \multicolumn{2}{c}{\textbf{Structure Guidance}}
& \multicolumn{2}{c}{\textbf{Semantic Guidance}}
& \multicolumn{2}{c}{\textbf{ACDC}}
& \multicolumn{2}{c}{\textbf{BTCV}}
\\

\cmidrule(lr){1-2}
\cmidrule(lr){3-4}
\cmidrule(lr){5-6}
\cmidrule(lr){7-8}
\cmidrule(lr){9-10}

ControlNet-Style
& Tri-Stream
& Residual
& SR Adapter
& Linear
& SA Refiner
& SSIM
& PSNR
& SSIM
& PSNR
\\
\midrule

\checkmark & --
& -- & --
& -- & --
& 84.2 & 26.9
& 79.4 & 26.8
\\

-- & \checkmark
& \checkmark & --
& -- & --
& 86.7 & 28.9
& 79.8 & 27.6
\\

-- & \checkmark
& -- & \checkmark
& -- & --
& 88.6 & 30.4
& 80.4 & 28.7
\\

-- & \checkmark
& -- & \checkmark
& \checkmark & --
& 89.7 & 32.7
& 81.9 & 29.3
\\

-- & \checkmark
& -- & \checkmark
& -- & \checkmark
& \textbf{91.2} & \textbf{34.4}
& \textbf{83.4} & \textbf{31.0}
\\

\bottomrule
\end{tabular}
\end{table*}

To comprehensively evaluate the performance of MedDiT4SR across diverse medical imaging modalities, we compare it with a wide range of representative medical image super-resolution methods on five benchmark datasets. The compared methods include modality-specific approaches, such as DiTMSR~\cite{tu2025taming}, PromptMR~\cite{xin2023kspacerefineimageprompting}, FIGCT~\cite{mo2026foveatedimaginggeometryctarchitecture}, CT-SRGAN~\cite{You_2020}, Fundus-GAN~\cite{zhao2024perception}, HistoSR~\cite{duan2024efficientdualbranchframeworkimplicit}, DMUISR~\cite{liu2024super}, as well as general-purpose restoration models, including MedSRGAN~\cite{gu2020medsrgan}, SG-SRNet~\cite{lu2025information}, and the recent diffusion transformer-based SD3-ControlNet.


From Table~\ref{tab:multi}, MedDiT4SR achieves the best or competitive performance across the five benchmark datasets, demonstrating that the proposed architecture can be effectively adapted to medical image super-resolution tasks from diverse imaging modalities. Compared with modality-specific super-resolution methods, MedDiT4SR delivers strong reconstruction performance on most datasets and evaluation metrics, suggesting the broad applicability of the shared architectural design under modality-specific training. The visual comparisons in Figure~\ref{fig:vis} further support the quantitative results, showing that MedDiT4SR reconstructs finer anatomical structures and visually more consistent textures than the competing methods, particularly in the highlighted regions.

\subsubsection{Domain Generalization}

Medical images acquired from different scanners and imaging protocols often exhibit substantial domain shifts. To evaluate cross-domain generalization, we train MedDiT4SR on a source dataset and directly evaluate it on unseen target datasets from the same modality without target-domain adaptation.

As shown in Table~\ref{tab:cross_dataset}, MedDiT4SR consistently achieves the best cross-dataset performance across different imaging modalities. Compared with existing medical SR methods and the large-scale diffusion baseline SD3-ControlNet, MedDiT4SR obtains higher PSNR scores on all evaluated target datasets under both $\times2$ and $\times4$ settings. The consistent improvements on breast ultrasound and cardiac MRI demonstrate that the proposed framework effectively transfers structural and semantic priors learned from the source domain to unseen target domains. These results indicate that MedDiT4SR exhibits strong robustness to domain shifts, making it well suited for real-world clinical applications.

\subsubsection{Ablation Study}
We conduct a comprehensive ablation study to verify the effectiveness of the proposed components. The results are summarized in Table~\ref{tab:ablation}. Replacing the ControlNet-style attention with the proposed Tri-Stream design, together with a simple residual connection, already yields consistent improvements, demonstrating the benefit of jointly modeling the LR, Noise, and Text Streams. The proposed SR Adapter further outperforms the residual design on both datasets by providing more effective structure-aware guidance. For semantic guidance, adding a linear modulation branch further improves reconstruction performance, while replacing it with the proposed SA Refiner achieves the best results. These consistent gains validate the effectiveness and complementary roles of Tri-Stream attention, the SR Adapter, and the SA Refiner in our proposed MedDiT4SR.

\subsubsection{Data Efficiency Evaluation}












\begin{table}[t]
\centering
\small
\setlength{\tabcolsep}{4pt}
\caption{Data-efficiency evaluation on the ACDC dataset under different training data ratios for $\times4$ super-resolution.}
\label{tab:data_efficiency}
\begin{tabular}{cccc}
\toprule
\textbf{Methods}
& \textbf{10\% Data}
& \textbf{50\% Data}
& \textbf{100\% Data} \\
\midrule

DiTMSR
& 82.0 $\pm$ 0.034
& 86.3 $\pm$ 0.032
& 90.4 $\pm$ 0.033\\

FIGCT
& 79.4 $\pm$ 0.031
& 82.7 $\pm$ 0.034
& 85.2 $\pm$ 0.030\\

Fundus-GAN
& 81.6 $\pm$ 0.030
& 84.8 $\pm$ 0.031
& 88.3 $\pm$ 0.026\\

HistoSR
& 78.2 $\pm$ 0.033
& 81.7 $\pm$ 0.035
& 84.9 $\pm$ 0.028\\

MedSRGAN
& 82.9 $\pm$ 0.030
& 85.7 $\pm$ 0.034
& 89.4 $\pm$ 0.028\\

SD3-ControlNet
& 80.5 $\pm$ 0.028
& 84.2 $\pm$ 0.030
& 88.6 $\pm$ 0.026 \\

\midrule

\textbf{MedDiT4SR}
& \textbf{86.3 $\pm$ 0.029}
& \textbf{90.1 $\pm$ 0.031}
& \textbf{91.2 $\pm$ 0.026} \\

\bottomrule
\end{tabular}
\end{table}

We evaluate MedDiT4SR under varying amounts of training data to assess its data efficiency and robustness under limited supervision. Specifically, models are trained using 10\%, 50\%, and 100\% of the ACDC training set.
We report the mean SSIM together with the corresponding standard deviation over multiple runs.

As shown in Table~\ref{tab:data_efficiency}, MedDiT4SR consistently achieves the best performance across all training data ratios. Notably, with only 10\% of the training data, it outperforms the strongest competing baseline by 3.4 percentage points, demonstrating strong data efficiency under limited supervision. Its advantage is maintained as the amount of training data increases. Moreover, MedDiT4SR exhibits small standard deviations across all settings, indicating stable and reliable performance over repeated runs. 
These results demonstrate that MedDiT4SR effectively leverages large-scale pre-trained diffusion priors, achieving strong data efficiency and reliable generalization under limited training data.

\subsubsection{Text Prompt Design}

\begin{table}[t]
\centering
\small
\setlength{\tabcolsep}{8pt}
\caption{Effect of text prompt design on BUSI dataset.}
\label{tab:prompt_design}
\begin{tabular}{ccccc}
\toprule
\multirow{2}{*}{\textbf{Prompt Design}}
& \multicolumn{2}{c}{$\mathbf{\times2}$ \textbf{Scale}}
& \multicolumn{2}{c}{$\mathbf{\times4}$ \textbf{Scale}} \\
\cmidrule(lr){2-3}
\cmidrule(lr){4-5}
& SSIM & PSNR
& SSIM & PSNR \\

\midrule

Contradictory
& 87.9 & 29.1
& 84.7 &  27.0\\

Missing Location
& 88.2 & 29.6
& 86.9 & 28.1\\

Overdescriptive
& 90.2 & 30.8
& 86.3 & 28.0 \\

Underdescriptive
& 89.9 & 30.7
& 87.0 & 28.3 \\

Empty Prompt
& 88.0 & 29.5
& 86.7 & 27.8 \\

\midrule

\textbf{Original}
& \textbf{90.6} & \textbf{31.3}
& \textbf{87.2} & \textbf{29.7} \\

\bottomrule
\end{tabular}
\end{table}

In clinical practice, textual descriptions accompanying medical images may be incomplete, ambiguous, or noisy due to variations in documentation styles and protocols. Therefore, a multimodal MedSR model should remain robust to imperfect or unavailable textual guidance.

As shown in Table~\ref{tab:prompt_design}, the original prompt consistently achieves the best performance under both the $\times2$ and $\times4$ settings. Removing spatial information, providing overly detailed or insufficient descriptions, and using an empty prompt lead to varying degrees of performance degradation, although MedDiT4SR retains competitive reconstruction quality in these cases. In contrast, contradictory prompts result in the most pronounced performance drop, reducing both SSIM and PSNR at both scales. This observation indicates that semantically consistent textual guidance is important for accurate reconstruction. Meanwhile, the relatively stable performance under incomplete, overdescriptive, underdescriptive, and empty prompts suggests that MedDiT4SR does not depend exclusively on perfectly matched textual inputs and can primarily rely on the LR image evidence when textual information is limited or unavailable.



\subsubsection{Downstream Task Evaluation}

\begin{table}[t]
    \centering
    \caption{
    Downstream segmentation performance on ACDC and BUSI using
    images reconstructed by different methods.
    }
    \label{tab:segmentation}
    \small
    \setlength{\tabcolsep}{2pt}
    \renewcommand{\arraystretch}{1.0}

    \begin{tabular}{cc*{4}{c}}
        \toprule
        \multirow{2}{*}{\textbf{Scale}}
        & \multirow{2}{*}{\textbf{Methods}}
        & \multicolumn{2}{c}{\textbf{ACDC}}
        & \multicolumn{2}{c}{\textbf{BUSI}} \\
        \cmidrule(lr){3-4}
        \cmidrule(lr){5-6}
        &
        & \textbf{Dice (\%)}
        & \textbf{IoU (\%)}
        & \textbf{Dice (\%)}
        & \textbf{IoU (\%)} \\
        \midrule

        \multirow{6}{*}{$\times2$}
        & DiTMSR
        & 86.14 & 78.33
        & 87.19 & 77.51 \\

        & MedSRGAN
        & 87.92 & 79.28
        & 88.69 & 81.84 \\

        & SG-SRNet
        & 85.67 & 76.80
        & 85.56 & 76.95 \\

        & SD3-ControlNet
        & 87.40 & 77.62
        & 88.13 & 80.79 \\

        & \textbf{MedDiT4SR}
        & \textbf{89.12}
        & \textbf{81.77}
        & \textbf{90.74}
        & \textbf{83.47} \\

        \cmidrule(lr){2-6}
        & GT
        & 90.29 & 82.35
        & 91.35 & 84.29 \\

        \midrule

        \multirow{6}{*}{$\times4$}
        & DiTMSR
        & 84.75 & 75.13
        & 84.96 & 75.88 \\

        & MedSRGAN
        & 87.37 & 78.65
        & 87.91 & 80.05 \\

        & SG-SRNet
        & 84.38 & 75.16
        & 84.79 & 74.82 \\

        & SD3-ControlNet
        & 86.40 & 76.07
        & 87.40 & 77.96 \\

        & \textbf{MedDiT4SR}
        & \textbf{88.96}
        & \textbf{81.25}
        & \textbf{90.02}
        & \textbf{82.83} \\

        \cmidrule(lr){2-6}
        & GT
        & 90.29 & 82.35
        & 91.35 & 84.29 \\

        \bottomrule
    \end{tabular}
\end{table}

To provide a comprehensive evaluation of the super-resolution results generated by different methods, we conduct downstream segmentation analyses using nnU-Net~\cite{isensee2021nnu} on the ACDC and BUSI datasets. Beyond perceptual quality, these task-based evaluations assess whether the reconstructed images preserve clinically relevant anatomical and lesion structures across different imaging modalities.

As shown in Table~\ref{tab:segmentation}, MedDiT4SR consistently achieves the best segmentation performance among all reconstruction methods on both datasets and at both upsampling scales, while remaining closest to the ground-truth results. Its advantage is maintained under the more challenging $\times4$ setting. These results indicate that MedDiT4SR better preserves clinically relevant anatomical and lesion structures.


\section{Conclusion}

In this paper, we present MedDiT4SR, the first framework to adapt a large-scale pre-trained diffusion transformer to medical image super-resolution. By jointly modeling the LR, Noise, and Text Streams, together with the SR Adapter and SA Refiner, MedDiT4SR integrates structural and semantic guidance throughout the denoising process. Experiments under both in-domain and within-modality cross-dataset settings demonstrate the effectiveness of adapting large-scale pre-trained DiT models to medical image super-resolution across diverse imaging domains.




\bibliography{aaai2027}


\end{document}